\documentstyle[11pt]{article}
\begin{document}

\title{\Large\bf
On the mass relation of a meson nonet} \vspace{1cm}
\author{ \small De-Min Li$^{1,2}$\footnote{E-mail: lidm@mail.ihep.ac.cn/lidm@zzu.edu.cn},
~~Hong Yu$^{2}$\footnote{E-mail: yuh@mail.ihep.ac.cn},~
Qi-Xing Shen$^{2}$\footnote{E-mail: shenqx@mail.ihep.ac.cn}\\
\small $^1$ Department of Physics, Zhengzhou University,
Zhengzhou, Henan 450052, P. R. China \\
\small$^2$Institute of High Energy Physics, Chinese Academy of Sciences,\\
\small P.O.Box $918~(4)$, Beijing 100039, P. R. China\\         }
\date{\today}
\maketitle
\vspace*{0.3cm}

\begin{abstract}
It is pointed out that the omission of the effects of the
transition between quarkonia or the assumption that the transition
between quarkonia is flavor-independent would result in the
inconsistent results for the pseudoscalar meson nonet. It is emphasized that
the mass relation of the non-ideal mixing meson nonets should incorporate the
effects of the flavor-dependent transition between quarkonia. The
new mass relations of a meson nonet are presented.
\end{abstract}

\vspace{1.5cm}


\newpage

\baselineskip 24pt

\section*{I. Introduction}
\indent

In the framework of the quark model, a $q\overline{q}$ meson nonet
contains two isoscalar states $\eta_8$ and $\eta_1$. In general,
these two isoscalar states can mix, which results in two physical
states $\eta$ and $\eta^\prime$.
\begin{equation}
\left(\begin{array}{c}
\eta\\
\eta^\prime
\end{array}\right)=U\left(\begin{array}{c}
\eta_8\\
\eta_1
\end{array}\right)=
\left(\begin{array}{cc}
\cos\theta&-\sin\theta\\
\sin\theta&\cos\theta
\end{array}\right)\left(\begin{array}{c}
\eta_8\\
\eta_1
\end{array}\right).
\label{angle}
\end{equation}
In the
$\eta_8=(u\overline{u}+d\overline{d}-2s\overline{s})/\sqrt{6}$ and
$\eta_1=(u\overline{u}+d\overline{d}+s\overline{s})/\sqrt{3}$
basis, the mass-squared matrix $M^2$ describing the mixing of
$\eta_8-\eta_1$ can be written as\cite{feclose}
\begin{equation}
M^2=\left(\begin{array}{cc}
M^2_{8}&M^2_{18}\\
M^2_{18}&M^2_{1}
\end{array}\right),
\label{feclose}
\end{equation}
which satisfies
\begin{equation}
UM^2U^{-1}=\left(\begin{array}{cc}
m^2_\eta&0\\
0&m^2_{\eta^\prime}
\end{array}\right),
\label{diag}
\end{equation}
From Eqs. (\ref{feclose}) and (\ref{diag}), one can obtain
\begin{eqnarray}
M^2_8+M^2_1=m^2_\eta+m^2_{\eta^\prime}.
\end{eqnarray}
Some authors believe that $M^2_8=m^2_8$ and $M^2_1=m^2_1$\cite{bura1,bura2}, where $m_8$ and $m_1$ are the
masses of the bare (before mixed) states $\eta_8$ and
$\eta_1$, respectively. Then based on the original Gell-Mann-Okubo
mass formula\cite{gell}
\begin{eqnarray}
m^2_8=\frac{1}{3}(4m^2_K-m^2_{\pi^0}),~~m^2_1=\frac{1}{3}(2m^2_K+m^2_{\pi^0}),
\label{gmo}
\end{eqnarray}
it is argued\cite{bura1}
that the mass spectrum of a meson nonet is linear:
\begin{eqnarray}
M^2_8+M^2_1=m^2_\eta+m^2_{\eta^\prime}=2m^2_K.
\label{err1}
\end{eqnarray}
However, for the pseudoscalar nonet, $m^2_\eta+m^2_{\eta^\prime}=1.217$ (GeV)$^2$, but
$2m^2_K=0.492$ (GeV)$^2$, therefore, it is believed\cite{bura2} that the original
Gell-Mann-Okubo mass formula is invalid for the pseudoscalar
nonet. Based on Regge phenomenology, the
new version of Gell-Mann-Okubo mass formula is given
as\cite{bura2}
\begin{eqnarray}
2M^2_S-M^2_N=4m^2_K-3m^2_{\pi^0},
\label{gmor}
\end{eqnarray}
where $M^2_N=\cos^2\alpha~ m^2_{\eta}+\sin^2\alpha~
m^2_{\eta^\prime}$, $M^2_S=\sin^2\alpha~ m^2_\eta+\cos^2\alpha~
m^2_{\eta^\prime}$ and $\sin\alpha=\frac{\sqrt{2}\cos\theta+\sin\theta}{\sqrt{3}}$.

Here, we want to emphasize that $M^2_8$ and $M^2_1$, the diagonal
elements of the mass matrix $M^2$ describing
the mixing of $\eta_8-\eta_1$, are rather different from $m^2_8$ and
 $m^2_1$, the masses-squared of the bare octet $\eta_8$ and singlet
 $\eta_1$, respectively, which are employed in the original
 Gell-Mann-Okubo mass formula. The left-hand side and the right-hand side
  of Eq. (\ref{err1})
are not balance does not result from that the original Gell-Mann-Okubo
mass formula is invalid for the pseudoscalar nonet, but from that the
effects of the transition amplitudes $Am_{88}$ and $Am_{11}$ are
not considered, where $Am_{88}$ and $Am_{11}$ denote
the transition amplitudes
of $\eta_8\leftrightarrow\eta_8$ and $\eta_1\leftrightarrow\eta_1$, respectively.
Also, the so-called new version Gell-Mann-Okubo mass formula, Eq.
(\ref{gmor}), is in fact derived in the
presence of the flavor-independent transition between
quarkonia. The assumption
that the transition between quarkonia is flavor-independent
would give the inconsistent results for the pseudoscalar meson nonet\cite{li1}.
The main purpose of this work is to emphasize that the mass relation of the
non-ideal mixing nonets should incorporate
the effects of the flavor-dependent transition between quarkonia.

\section*{II. Effects of the transition amplitudes of $q\overline{q}\leftrightarrow q^\prime\overline{q^\prime}$}
\indent

In the $N=(u\overline{u}+d\overline{d})/\sqrt{2}$ and
$S=s\overline{s}$ basis, the general form of the mass-squared matrix
describing the
mixing of the physical states $\eta$ and $\eta^\prime$ can be written
as\cite{feclose}
\begin{equation}
{M^\prime}^2=\left(\begin{array}{cc}
M^2_N&A_{NS}\\
A_{NS}&M^2_S
 \end{array}\right),
 \label{matrix}
\end{equation}
with
\begin{eqnarray}
M^2_N=m^2_N+A_{NN},~~M^2_S=m^2_S+A_{SS},
\label{mns}
\end{eqnarray}
where $m_N$ and $m_S$ are the masses of the bare states $N$ and
$S$, respectively; $A_{NN}$, $A_{SS}$ and $A_{SN}$ are the
transition amplitudes of $ N\leftrightarrow N$,
$S\leftrightarrow S$ and $N\leftrightarrow S$, respectively.
If the transition between  quarkonia is
flavor-dependent\cite{flavor,close,chanwitz}, $A_{NN}$, $A_{SS}$ and $A_{SN}$ usually are
parameterized as\cite{li1,Turnau,kawai}
\begin{eqnarray}
A_{SS}=A,~~A_{NN}=r^2A,~~A_{SN}=rA.
\label{parameter}
\end{eqnarray}
$r^2=2$, i.e., $A_{SS}=A$, $A_{NN}=2A$ and $A_{SN}=\sqrt{2}A$ means that the transition between quarkonia is
flavor-independent\cite{schnitzer,rosner}.
Owing to
\begin{equation}
(N,S)=(\eta_8,\eta_1)R=(\eta_8,\eta_1)\left(\begin{array}{cc}
\sqrt{\frac{1}{3}}&-\sqrt{\frac{2}{3}}\\
\sqrt{\frac{2}{3}}&\sqrt{\frac{1}{3}}
\end{array}\right),
\label{nangle}
\end{equation}
the mass-squared matrices $M^2$ and ${M^\prime}^2$ can be connected
by
\begin{equation}
M^2=R{M^\prime}^2R^{-1}.
\label{connect}
\end{equation}

$N$ is the orthogonal partner of $\pi^0$, the isovector state of a meson nonet, and one can expect that $N$ degenerates with
$\pi^0$ in effective quark masses, therefore we can assume
$m^2_N=m^2_{\pi^0}$, which is a widely adopted assumption\cite{kawai,feldmann,bura3}.
According to the other form of the original Gell-Mann-Okubo mass
formula\cite{gell}
$m^2_N+m^2_S=2m^2_K$, one can get
\begin{eqnarray}
&&M^2_8=\frac{1}{3}(4m^2_K-m^2_{\pi^0})+
\frac{1}{3}Ar^2-\frac{2\sqrt{2}}{3}Ar+\frac{2}{3}A,
\label{element1}
\\
&&M^2_{18}=-\frac{2\sqrt{2}}{3}(m^2_K-m^2_{\pi^0})
+\frac{\sqrt{2}}{3}Ar^2-\frac{1}{3}Ar-
\frac{\sqrt{2}}{3}A,
\label{element2}
\\
&&M^2_1=\frac{1}{3}(2m^2_K+m^2_{\pi^0})
+\frac{2}{3}Ar^2+\frac{2\sqrt{2}}{3}Ar+\frac{1}{3}A,
\label{element3}
\end{eqnarray}
where
\begin{equation}
A=
\frac{(m^2_{\eta^\prime}-2m^2_K+m^2_{\pi^0})(m^2_{\eta}-2m^2_K+m^2_{\pi^0})}
{2(m^2_{\pi^0}-m^2_K)},
\end{equation}
\begin{equation}
r^2=\frac{(m^2_{\eta}-m^2_{\pi^0})(m^2_{\pi^0}-m^2_{\eta^\prime})}
{(m^2_{\eta^\prime}-2m^2_K+m^2_{\pi^0})(m^2_{\eta}-2m^2_K+m^2_{\pi^0})}.
\label{rr}
\end{equation}
Comparing  Eq. (\ref{gmo}) with Eqs. (\ref{element1})$\sim$
(\ref{element3}), one can have
\begin{eqnarray}
&&M^2_{8}=m^2_8+Am_{88},\\
&&M^2_{1}=m^2_1+Am_{11},\\
&&Am_{88}=\frac{1}{3}Ar^2-\frac{2\sqrt{2}}{3}Ar+\frac{2}{3}A,\\
&&Am_{11}=\frac{2}{3}Ar^2+\frac{2\sqrt{2}}{3}Ar+\frac{1}{3}A,
\end{eqnarray}
which shows that $M^2_8$ and $M^2_1$ are rather different from
$m^2_8$ and $m^2_1$.

Based on the above relations, the mass relation of a meson nonet can be
read as
\begin{eqnarray}
M^2_8+M^2_1=m^2_\eta+m^2_{\eta^\prime}=2m^2_K+A(r^2+1).
\label{nm81}
\end{eqnarray}
If $A$ is set to be zero, Eq. (\ref{nm81}) can be reduced to Eq.
(\ref{err1}), Therefore, it is
in the absence of the effects of
the transition amplitudes $Am_{88}$ and $Am_{11}$ that the
argument\cite{bura1} $m^2_\eta+m^2_{\eta^\prime}=2m^2_K$ is given.

Both Eq. (\ref{err1}) and Eq. (\ref{nm81}) are deduced from the original
Gell-Mann-Okubo mass formula, the only difference between Eqs. (\ref{err1}) and
(\ref{nm81}) is that Eq. (\ref{nm81}) incorporates the effects of
the transition amplitudes $Am_{88}$ and $Am_{11}$. However, for the pseudoscalar nonet
 Eq. (\ref{err1})
 is obviously invalid while Eq.
(\ref{nm81}) formula does hold with a high accuracy (both sides of Eq. (\ref{nm81})
are equal to $1.217$ (GeV)$^2$), which indicates that Eq. (\ref{err1})
is invalid for the pseudoscalar nonet does not result from that the original
Gell-Mann-Okubo
mass formula is invalid for the pseudoscalar nonet, but from the
omission of the effects of
the transition amplitudes $Am_{88}$ and $Am_{11}$.

\section*{III. Inconsistency from the flavor-independent transition amplitudes of $q\overline{q}\leftrightarrow q^\prime\overline{q^\prime}$}
\indent

Now we turn to discuss why the so-called new version of Gell-Mann-Okubo mass
formula Eq. (\ref{gmor}) is in fact derived from the assumption
that the transition between quarkonia is
flavor-independent.

From Eqs. (\ref{diag}), (\ref{matrix}) and (\ref{connect}), we
have
\begin{eqnarray}
M^2_N+M^2_S=m^2_\eta+m^2_{\eta^\prime},
\label{trace}
\end{eqnarray}
then from Eqs. (\ref{gmor}) and (\ref{trace}), the following
equations can be given
\begin{eqnarray}
M^2_N=\frac{2m^2_\eta+2m^2_{\eta^\prime}+3m^2_{\pi^0}-4m^2_K}{3},~~M^2_S=\frac{4m^2_K-3m^2_{\pi^0}+m^2_\eta+m^2_{\eta^\prime}}{3}.
\label{nmns}
\end{eqnarray}
With $m^2_N+m^2_S=2m^2_K$ and $m^2_N=m^2_{\pi^0}$, from Eqs. (\ref{mns}) and (\ref{nmns}),
$A_{NN}$ and $A_{SS}$ can be read as
\begin{eqnarray}
&&A_{NN}=M^2_N-m^2_N=\frac{2m^2_\eta+2m^2_{\eta^\prime}-4m^2_K}{3},\\
&&A_{SS}=M^2_S-m^2_S=\frac{m^2_\eta+m^2_{\eta^\prime}-2m^2_K}{3}.
\end{eqnarray}
So $A_{NN}$ and $A_{SS}$ satisfy
\begin{eqnarray}
A_{NN}=2A_{SS},
\label{independent}
\end{eqnarray}
which implies that the transition between
quarkonia is flavor-independent, and one can deduce that
\begin{eqnarray}
A_{SN}=\sqrt{2}A_{SS}.
\label{independent1}
\end{eqnarray}

We note that with the assumption $m^2_N=m^2_{\pi^0}$, the original Gell-Mann-Okubo mass formula
$m^2_S+m^2_N=2m^2_K$ can be re-expressed by
\begin{eqnarray}
2m^2_S-m^2_N=4m^2_K-3m^2_{\pi^0}.
\label{reexpressed}
\end{eqnarray}
From  Eqs. (\ref{mns}), (\ref{independent}) and (\ref{reexpressed}), we can have
\begin{eqnarray}
2M^2_S-M^2_N\equiv 2m^2_S-m^2_N=4m^2_K-3m^2_{\pi^0},
\end{eqnarray}
which shows that the implicit assumption $A_{NN}=2A_{SS}$ exists in
Eq. (\ref{gmor}).

In fact, based on $M^2_N=\cos^2\alpha~ m^2_{\eta}+\sin^2\alpha~
m^2_{\eta^\prime}$, $M^2_S=\sin^2\alpha~ m^2_\eta+\cos^2\alpha~
m^2_{\eta^\prime}$ and
$\sin\alpha=\frac{\sqrt{2}\cos\theta+\sin\theta}{\sqrt{3}}$,
the following relation can be given
\begin{eqnarray}
2M^2_S-M^2_N\equiv\cos^2\theta~m^2_\eta+\sin^2\theta~m^2_{\eta^\prime}+\sqrt{2}
 \sin 2\theta(m^2_\eta-m^2_{\eta^\prime}).
 \label{main1}
 \end{eqnarray}
From Eqs. (\ref{angle}), (\ref{feclose}) and (\ref{diag}), one can have
 \begin{eqnarray}
\cos^2\theta=\frac{m^2_{\eta^\prime}-M^2_8}{m^2_{\eta^\prime}-m^2_\eta},~~
\sin 2\theta=\frac{2M^2_{18}}{m^2_{\eta^\prime}-m^2_\eta}.
\label{cossin}
 \end{eqnarray}
According to Eqs. (\ref{element1}), (\ref{element2}),  (\ref{main1}) and (\ref{cossin}), one can have
 \begin{eqnarray}
 2M^2_S-M^2_N\equiv\cos^2\theta~m^2_\eta+\sin^2\theta~m^2_{\eta^\prime}+\sqrt{2}
 \sin 2\theta(m^2_\eta-m^2_{\eta^\prime})
 =4m^2_K-3m^2_{\pi^0}-A(r^2-2).
 \label{main}
 \end{eqnarray}
 Obviously, if $r^2$ is set to be 2, Eq. (\ref{main}) can be reduced
 to Eq. (\ref{gmor}).
Therefore, it is in the presence of the assumption that the transition between
quarkonia is flavor-independent, that Eq. (\ref{gmor}) is derived.

Under the assumption that
the transition between quarkonia is flavor-independent,
i.e., from Eqs. (\ref{independent}) and (\ref{independent1}),
the mass matrix Eq. (\ref{matrix}) can be reduced to
\begin{equation}
{M^{\prime\prime}}^2=\left(\begin{array}{cc}
m^2_N+2A&\sqrt{2}A\\
\sqrt{2}A&m^2_S+A
 \end{array}\right),
 \label{nmatrix}
\end{equation}
where $A\equiv A_{SS}$.
Diagonalizing the above mass matrix, one can have
\begin{eqnarray}
&&m^2_N+m^2_S+3A=m^2_\eta+m^2_{\eta^\prime},\\
&&(m^2_N+2A)(m^2_S+A)-2A^2=m^2_\eta m^2_{\eta^\prime},
\end{eqnarray}
By eliminating $A$ from the above two relations, one can get
Schwinger's original nonet mass formula\cite{schwinger}
\begin{equation}
(4m^2_K-3m^2_{\eta}-m^2_{\pi^0})(3m^2_{\eta^\prime}+m^2_{\pi^0}-4m^2_K)
=8(m^2_K-m^2_{\pi^0})^2,
\label{schwinger}
\end{equation}
here we still use $m^2_N+m^2_S=2m^2_K$ and the assumption
$m^2_N=m^2_{\pi^0}$.

For the pseudoscalar nonet, the left-hand side of Eq.
(\ref{schwinger}) is 0.1178 (GeV)$^4$, while the right-hand side of Eq. (\ref{schwinger})
is 0.4140 (GeV)$^4$, which clearly shows that for the pesudoscalar meson nonet, the assumption that
the transition
between quarkonia is flavor-independent can result in the inconsistent results.

However, if we assume that the transition
between quarkonia is flavor-dependent, i.e., we
use the mass matrix $\left(\begin{array}{cc}
m^2_N+r^2A&rA\\
rA&m^2_S+A
 \end{array}\right)$ to
describing the mixing of $\eta$ and $\eta^\prime$, in the presence of $m^2_N+m^2_S=2m^2_K$ and
$m^2_N=m^2_{\pi^0}$, the new
version of Schwinger's nonet mass formula can be read as
\begin{equation}
[2r^2m^2_K-(1+r^2)m^2_{\eta}-(r^2-1)m^2_{\pi^0}]
[(1+r^2)m^2_{\eta^\prime}+(r^2-1)m^2_{\pi^0}-2r^2m^2_K]=
4r^2(m^2_K-m^2_{\pi^0})^2.
\label{newsch}
\end{equation}
For the pseudoscalar nonet, both sides of Eq. (\ref{newsch}) are
equal to 0.6788 (GeV)$^4$.

The original Gell-Mann-Okubo mass formula and the assumption $m^2_N=m^2_{\pi^0}$ are
incorporated in both Eq. (\ref{schwinger}) and Eq. (\ref{newsch}), however,
for the pseudoscalar nonet,  Eq. (\ref{schwinger}) is obviously invalid,
while Eq. (\ref{newsch}) holds with a
high accuracy, which indicates that for the pseudoscalar meson nonet,
the assumption that the transition between quarkonia is
flavor-independent can result in the inconsistent results. The
correct mass relation of the pseudoscalar meson nonet should
include the effects of the flavor-dependent transition between
quarkonia.

\section*{IV. Concluding remarks}
\indent

Generally speaking, the mixing
of the isoscalar states of a meson nonet results from some extra interactions
which not only cause
the breaking of SU(3) flavor symmetry
but also lead to the transition between quarkonia\cite{feclose,Turnau,kawai,isgur,fuchs}.
 As pointed out by Ref.\cite{isgur}, in the absence of the
 transition between quarkonia, the breaking of SU(3) flavor
 symmetry only results in the ideal mixing of the isoscalar states
 of a meson nonet. That is to say, for the ideal mixing meson nonet, the
 effects of the transition between quarkonia can be ignored,
 however, for the non-ideal mixing meson nonet, the effects of
 the transition between quarkonia should be rather important. Therefore, for the
 almost ideal mixing meson nonets, the effects of the flavor-independent transition
 between quarkonia and those of the
 flavor-dependent transition between quarkonia can be ignored.
 In fact, the results of Refs.\cite{bura1,bura2,li1,li2} all indicate
 that for the almost ideal mixing meson nonets such as the vector, tensor and axial vector
 meson nonets,  both Eq. (\ref{err1}) based on the omission of the transition
 amplitudes and Eq. (\ref{schwinger}) based on the assumption that the
 transition is flavor-independent can hold with a higher accuracy. However, for
 the pseudoscalar nonet which is the non-ideal mixing meson nonet,
 both Eq. (\ref{err1}) and Eq. (\ref{schwinger}) are obviously invalid while
 both Eq. (\ref{nm81}) and
  Eq. (\ref{newsch}) which all incorporate the effects of the flavor-dependent
  transition between quarkonia hold with rather high accuracy, which shows that
  the correct mass relation of
   the non-ideal mixing meson nonets should include the effects of the flavor-dependent transition.
Therefore, the mass relations such as Eqs. (\ref{nm81}),
(\ref{main}) and (\ref{newsch}) which include the effects of the
flavor-dependent transition between quarkonia hold not only for
the non-ideal mixing meson nonets but also for the almost ideal
mixing meson nonets.

In conclusion, we point out that the invalidity of Eq.
(\ref{err1}) for the pseudoscalar meson nonet arises from the omission of the
effects of the
transition between quarkonia, and that the so-called new version
of Gell-Mann-Okubo mass formula in fact is based on the assumption
that the transition between quarkonia is flavor-independent which
can result in the inconsistent results for the pseudoscalar meson nonet.
We emphasize that the correct mass relation of the non-ideal mixing meson nonets
 should include the effects of the flavor-dependent transition between quarkonia.
The new mass relations such as Eqs. (\ref{nm81}),
(\ref{main}) and (\ref{newsch}) which include the effects of the
flavor-dependent transition between quarkonia hold for all the meson nonets.

\section*{Acknowledgments}
This project is supported by the National Natural Science
Foundation of China under Grant Nos. 19991487, 19835060 and 10047003.


\begin{thebibliography}{99}
\bibitem{feclose} F. E. Close, An introduction to quarks and
partons (London, Academic Press, 1979 )
\bibitem{bura1} L. Burakovsky and L. P. Horwitz, Found. Phys.
Lett. 9 (1996) 561; hep-ph/9608301.
\bibitem{bura2} L. Burakovsky and T. Goldman, hep-ph/9708498.
\bibitem{gell} S. Okubo, Prog. Theor. Phys. 27 (1962) 949.
\bibitem{li1} De-Min Li, Hong Yu and Qi-Xing Shen, Chin. Phys.
Lett. 18 (2001) 184.
\bibitem{flavor} A. De Rujula, H. Georgi and S. Glashow, Phys.
Rev. D 12 (1975) 147.
\bibitem{close} F. E. Close, Prog. Theor. Phys. 51 (1988) 833.
\bibitem{chanwitz} M. Chanowitz and S. R. Sharpe, Phys. Lett. B
132 (1983) 413.
\bibitem{Turnau} J. Turnau, Z. Phys. C 23 (1984) 89.
\bibitem{kawai} E. Kawai, Phys. Lett. B 124 (1983) 262.
\bibitem{schnitzer} H. J. Schnitzer, Nucl. Phys. B 207 (1982) 131.
\bibitem{rosner} J. L. Rosner, Phys. Rev. D 27 (1983) 1101.
\bibitem{feldmann} T. Feldmann, P. Kroll and B. Stech, Phys. Rev.
D 58 (1998) 114006.
\bibitem{bura3} L. Burakovsky and P. R. Page, Eur. Phys. J. C 12
(2000) 489.
\bibitem{schwinger}J. Schwinger, Phys. Rev. Lett. 12 (1964) 237.
\bibitem{isgur}N. Isgur, Phys. Rev. D 12 (1975) 3770.
\bibitem{fuchs} N. H. Fuchs, Phys. Rev. D 14 (1976) 1912.
\bibitem{li2} De-Min Li, Hong Yu and Qi-Xing Shen, Chin. Phys.
Lett. 17 (2000) 558.
\end{thebibliography}
\end{document}